\documentclass[preprint,superscriptaddress,eqsecnum,nofootinbib]{revtex4}
\usepackage{hyperref}
\usepackage{graphicx}
\usepackage{epsfig}
\usepackage{slashed}
\usepackage{MnSymbol}

\begin{document}

\newcommand{\be}{\begin{equation}}
\newcommand{\ee}{\end{equation}}
\newcommand{\bea}{\begin{eqnarray}}
\newcommand{\eea}{\end{eqnarray}}
\newcommand{\no}{\noindent}

\title{Path Integral for Off-Shell Supersymmetry}
\author{Chiu Man Ho} \email{cmho@msu.edu}
\affiliation{Department of Physics and Astronomy, Michigan State University, East Lansing, MI 48824, USA}

\date{\today}

\begin{abstract}
Off-shell supersymmetry, which restricts sparticles to appear only off-shell, solves the gauge hierarchy problem and unifies the
gauge couplings in the usual way. Without introducing any new interactions or exacerbating the naturalness, this idea has been proposed
to evade the direct detection bounds on sparticles. The original realization of off-shell supersymmetry relies on applying a generalized
quantization scheme solely to sparticles. In this paper, we formulate off-shell supersymmetry from the path-integral approach.

\end{abstract}

\maketitle

\section{Introduction}

Supersymmetry (SUSY) offers an elegant solution to the gauge hierarchy problem. In the Minimal Supersymmetric Standard Model (MSSM),
the three gauge couplings are unified at $M_{\rm GUT} \simeq  2 \times 10^{16}$ GeV, which is desirable for the Grand Unified Theory (GUT).
This makes suppersymmetry a promising idea. Since SUSY is most naturally realized at the TeV scale, we had expected to discover
sparticles in the first run of the Large Hadron Collider (LHC). However, no signal of sparticles has been observed so far and the hope
is now on the LHC Run II with a collider energy $13-14$ TeV. For an overview of sparticle search at the LHC, see \cite{SUSY_ATLAS, SUSY_CMS}.
The null search results for colored sparticles such as the gluino $\tilde g$ and squarks $\tilde q$ imply lower bounds on their masses with roughly $m_{\tilde g} \gtrsim 1$ TeV and $ m_{\tilde q} \gtrsim $ 600 GeV. The lower bounds on non-colored sparticles such as sleptons, charginos and neutralinos are not so severe, with roughly a few hundred GeV.

It is possible that sparticles might have actually been produced at the LHC but have escaped from the detection because
their signals are too difficult to be distinguished from the Standard Model (SM) background. This situation occurs when the energy of hadronic/leptonic jets and the missing transverse energy associated with the sparticle cascade decay are significantly reduced in some SUSY models, such as the so-called Compressed SUSY~\cite{Compressed}, Stealth SUSY~\cite{Stealth} and $R$-parity violating SUSY models~\cite{RpV}.
For these models, the lower bounds on sparticle masses can be relaxed up to a few hundred GeV. However, it is not easy to construct such models in a natural way~\cite{SUSYRev1,SUSYRev2}. These models cannot completely hide the direct signal of sparticles forever. With the substantially
improved sensitivity at the LHC Run II, it is conceivable that they will be probed and constrained soon.

Contrary to the SUSY models which can tentatively hide the sparticles from direct detection, we would like to provide an $R$-parity preserving
MSSM scenario that can survive even if sparticles may never be directly observed at the LHC. Our guiding principle is to retain simplicity
and naturalness as much as possible.

Supersymmetry at the Lagrangian does not dictate the way we should quantize the fields in the supermultiplets, and so we have the freedom
to quantize the SM particles and their superpartners differently. In \cite{me}, we proposed a generalized
quantization scheme under which a particle can only appear off-shell, while its contributions to quantum corrections behave exactly
the same as in the usual quantum field theory (QFT). We applied this quantization scheme solely to the sparticles in the $R$-parity preserving MSSM.
Thus sparticles can only appear off-shell.\footnote{After \cite{me} appeared on arXiv, we were informed that this
possibility was wished in \cite{Ghilencea}.} Without introducing any new interactions or exacerbating the naturalness, the sparticles could be light but would completely escape the direct detection at any experiments such as the LHC. However, our theory still retains the same desirable features of the usual MSSM at the quantum level. For instance, the gauge hierarchy problem is solved and the three MSSM gauge couplings are unified in the usual way. Although direct detection of sparticles is impossible, their existence can be revealed by precise measurements of some observables (such as the running QCD coupling) that may receive quantum corrections from them and have sizable deviations from the SM predictions. The experimental constraints from the indirect sparticle search are still applicable and the same as in the usual MSSM.

Of course, the lightest sparticle (e.g. neutralino) would no longer be a viable dark matter candidate if it can only appear off-shell. However,
this should not be considered as a serious deficiency of our MSSM scenario. Providing a viable dark matter candidate is a just bonus of the
usual MSSM, and it is easy enough to construct other models that give a promising dark matter candidate. Superpsymmetry is most crucial for
solving the gauge hierarchy problem and unification of the three MSSM gauge couplings.

In principle, one could apply our generalized quantization scheme to any other theories so as to evade the corresponding bounds from
direct detection. Nevertheless, sparticles are particularly well-suited for the generalized quantization. The reason is that the most
important merit of sparticles is due to their off-shell quantum contributions. For instance, we do not need on-shell sparticles in order to
solve the gauge hierarchy problem and unify the three MSSM gauge couplings.

The path-integral approach is perhaps the most consistent and convenient way to quantize the non-abelian gauge fields. It also facilitates the study of non-perturbative effects in quantum field theory. A consistent formulation from the path-integral approach makes a theory more promising. The original realization of off-shell supersymmetry has relied on applying the generalized quantization scheme solely to sparticles. In order to better justify its consistency, we formulate off-shell supersymmetry from the path-integral approach in this paper.

\section{Path-Integral Quantization in Quantum Field Theory}

In this section, we briefly review the path-integral quantization for scalars and fermions in the usual QFT.
We refer the readers to \cite{Ryder} for more details. In the next section, we will demonstrate how should these path-integrals
be modified in order to restrict scalars and fermions to appear only off-shell.

For a non-interacting real scalar $\phi$ with mass $m$, the generating functional in the presence of a source $J$ is given by
\bea
\label{generateS}
Z[J] = \int\, \mathcal{D}\phi \,\, \exp\left\{\,i\,\int\,d^4x\,\left(\,\,\mathcal{L}(\phi) + J(x)\,\phi(x)\,\,\right)  \,\right\}\,,
\eea
where $\mathcal{L}(\phi)$ is the free Lagrangian for the $\phi$ field. $Z[J]$ can be evaluated and normalized to give
\bea
\label{generateS0}
Z_0[J] = \exp\left(-\frac{i}{2}\,\int\,d^4x\,d^4y\,\, J(x) \, \Delta(x-y)\, J(y) \right)\,.
\eea
The Green's function $\Delta(x-y)$ obeys the inhomogeneous equation $\left(\,\Box^2 + m^2-i\epsilon\,\right)\,\Delta(x-y) = - \delta^{(4)}(x-y)$
and it reads
\bea
\Delta(x-y) = \int\,\frac{d^4 p}{(2\pi)^4}\,e^{-ip(x-y)} \,\frac{1}{p^2-m^2 +i\epsilon }\,.
\eea
The $n$-point correlation function can be obtained from
\bea
\label{correlateS}
\langle 0 | \, T \left\{ \,\phi(x_1) \,\cdots\, \phi(x_n)\,\right\} \,| 0 \rangle
= \left.\frac{1}{i^n}\,\frac{\delta^n\, Z_0[J]}{\delta J(x_1) \,\cdots\, \delta J(x_n)}\right|_{J=0}\,.
\eea
For a non-interacting complex scalar, $\mathcal{D}\phi$ in Eq. \eqref{generateS} is replaced by
$\mathcal{D}\phi^\dagger\,\mathcal{D}\phi$ and $J(x)\,\phi(x)$ by $J^\dagger(x)\,\phi(x) + \phi^\dagger(x)\,J(x) $.
In Eq. \eqref{generateS0}, $J(x)$ is replaced by $J^\dagger(x)$ and the overall factor of 1/2 is absent.
The $2n$-point correlation function is
\bea
\label{correlateComplexS}
\langle 0 | \, T \left\{ \,\phi(x_1)\,\phi^\dagger(x_2)\,\cdots\,\phi(x_{2n-1})\,\phi^\dagger(x_{2n})\,\right\} \,| 0 \rangle
= \left.\frac{1}{i^{2n}}\,\frac{\delta^{2n}\, Z_0[J, J^\dagger]}{\delta J^\dagger(x_1)\,J(x_2)\,\cdots\,J^\dagger(x_{2n-1})\,\delta J(x_{2n})}\right|_{J=J^\dagger=0}\,. \nonumber \\
\eea

For a non-interacting fermion $\psi$ with mass $m$, the generating functional in the presence of Grassmannian sources $\eta$
and $\bar\eta$ is given by
\bea
Z[\eta, \bar\eta] = \int\, \mathcal{D}\bar\psi \,\mathcal{D} \psi \,\, \exp\left\{\,i\,\int\,d^4x\,\left(\,\,\mathcal{L}(\bar\psi,\psi)
+ \bar\eta(x)\,\psi(x)+\bar\psi(x)\,\eta(x)\,\,\right)  \,\right\}\,,
\eea
where $\mathcal{L}(\bar\psi,\psi)$ is the free Lagrangian for the $\psi$ field. $Z[\eta, \bar\eta]$ can be evaluated and normalized to give
\bea
Z_0[\eta, \bar\eta] = \exp\left(-i\,\int\,d^4x\,d^4y\,\, \bar\eta(x) \, \mathcal{S}(x-y)\, \eta(y) \right)\,.
\eea
The Green's function $\mathcal{S}(x-y)$ obeys the inhomogeneous equation
$\left(\,\Box^2 + m^2-i\epsilon\,\right)\,\mathcal{S}(x-y) = - \delta^{(4)}(x-y)$ and it reads
\bea
\mathcal{S}(x-y) = \int\,\frac{d^4 p}{(2\pi)^4}\,e^{-ip(x-y)} \,\frac{\slashed{p}+m}{p^2-m^2 +i\epsilon } \,.
\eea
The $2n$-point correlation function can be obtained from
\bea
\langle 0 | \,T \left\{ \,\psi(x_1)\bar\psi(x_2)\,\cdots\,\psi(x_{2n-1})\bar\psi(x_{2n})\,\right\} \,| 0 \rangle  = \left. \frac{\delta^{2n}\,
Z_0[\eta, \bar\eta]}{\delta\bar\eta(x_{1})\,\delta\eta(x_{2}) \,\cdots\, \delta\bar\eta(x_{2n-1})\,\delta\eta(x_{2n})}\right|_{\eta=\bar\eta=0}\,.
~~~
\eea

When interactions are present, the most general generating functional involving a scalar $\phi$, a fermion $\psi$, an abelian vector boson
$A_\mu$ and a nonabelian vector boson $A^a_\nu$ is given by
\bea
Z[J, J^\dagger, \eta, \bar\eta, J_\mu, J^a_\nu] = \exp\left(\,i\,\int\, d^4x\,\,\mathcal{L}_I \,\right)\, Z_{0}\,,
\eea
where $Z_{0}$ is the product of the normalized generating functionals for the scalar, fermion, abelian and nonabelian
vector bosons. $\mathcal{L}_I$ is the Lagrangian that contains the interaction operators under consideration, but with the fields for the scalar, fermion, abelian and nonabelian vector bosons substituted by:
\bea
\phi(x) ~ &\rightarrow& ~ \frac{1}{i}\,\frac{\delta}{\delta J(x)}~, ~~~~~~\textrm{for\, real}~ \phi \\
\phi(x) ~ &\rightarrow& ~ \frac{1}{i}\,\frac{\delta}{\delta J^\dagger(x)}~, ~~~~~\textrm{for\, complex}~ \phi \\
\phi^\dagger(x) ~ &\rightarrow& ~ \frac{1}{i}\,\frac{\delta}{\delta J(x)}~, ~~~~~~\textrm{for\, complex}~ \phi \\
\psi(x) ~ &\rightarrow& ~ \frac{1}{i}\,\frac{\delta}{\delta \bar\eta}~, ~~~~~~~~
\bar\psi(x) ~ \rightarrow ~ -\frac{1}{i}\,\frac{\delta}{\delta \eta}~, \\
A_\mu(x) ~ &\rightarrow& ~ \frac{1}{i}\,\frac{\delta}{\delta J^\mu}~, ~~~~~~
A^a_\mu(x) ~ \rightarrow ~ \frac{1}{i}\,\frac{\delta}{\delta J_a^\mu}~.
\eea

\section{Path-Integral Approach to Off-Shell Supersymmetry}

In \cite{me}, we proposed a generalized quantization scheme which restricts particles to appear only off-shell. We applied this quantization scheme solely to the sparticles in the $R$-parity preserving MSSM. Thus sparticles can only appear off-shell. (In the MSSM, we do not have spin-1 sparticles, so the application of our generalized quantization scheme to vector bosons is not relevant.) Under the generalized quantization scheme,
we derived the propagators for scalars and fermions \cite{me}. The propagator for scalars is given by
\bea
\label{P_S}
\langle 0| \,T\{\,\phi(x)\,\phi(y)\,\}\,| 0 \rangle = \int\,\frac{d^4\,p}{(2\pi)^4}\,\, e^{-ip(x-y)}\,\,
\frac{1}{2}\,\left(\,\frac{i}{p^2-m^2+i\,\epsilon} + \frac{i}{p^2-m^2-i\,\epsilon} \,\right)\,,
\eea
while that for Dirac fermions is given by
\bea
\label{P_F}
&& \langle 0| \,T\{\,\psi_\alpha(x)\,\bar\psi_\beta(y)\,\}\,| 0 \rangle \,\equiv i\, \mathcal{S}_{\alpha\beta}(x-y) \nonumber  \\
&=& \int\,\frac{d^4 p}{(2\pi)^4}\, e^{-ip(x-y)} \,\left(\,\slashed{p}+m\,\right)_{\alpha\beta}\,\, \frac{1}{2}\,\left(\,\frac{i}{p^2-m^2+i\,\epsilon} + \frac{i}{p^2-m^2-i\,\epsilon} \,\right)\,.
\eea
For Majorana fermions with $\psi = \psi^{c} \equiv C\,\bar{\psi}^{T}$ where $C$ is the charge conjugation matrix, there are two additional propagators:
\bea
\langle 0| \,T\{\,\psi_\alpha(x)\,\psi_\beta(y)\,\}\,|0\rangle &=& \left[\,C^{-1}\,\mathcal{S}(x-y)\,\right]_{\alpha\beta}\,, \\
\langle 0| \,T\{\,\bar{\psi}_\alpha(x)\,\bar{\psi}_\beta(y)\,\}\,|0\rangle &=& \left[\,-\mathcal{S}(x-y)\,C\,\right]_{\alpha\beta}\,.
\eea

As shown in \cite{me}, the propagators for both scalars and fermions turn out to be half-retarded and half-advanced.
Under the generalized quantization scheme, a particle simultaneously propagates positive and negative energy modes both forward and backward in time with equal amplitudes. Therefore, the particle behaves like a ``standing wave" (in time) --- there is \emph{no} net energy flux being transferred. Since an on-shell particle carries a nonzero net energy flux, a particle propagated by the half-retarded and half-advanced propagator \emph{cannot} be on-shell.

In this section, we would like to demonstrate how should the usual path-integrals for scalars and fermions
be modified in order to restrict them to appear only off-shell. Under the generalized quantization scheme proposed in \cite{me},
a scalar or fermionic field, \emph{as a single field}, acquires both of the propagators $\frac{1}{p^2-m^2+i\,\epsilon}$ and
$\frac{1}{p^2-m^2-i\,\epsilon}$. This suggests that in the path-integral formulation, a single $J$ or a single pair of
$\eta$ and $\bar\eta$ is not sufficient. We need separate (Grassmannian) sources, which can account for each of
$\frac{1}{p^2-m^2+i\,\epsilon}$ and $\frac{1}{p^2-m^2-i\,\epsilon}$, for the same field. A natural way of doing this is to
introduce $\{J_+,\, J_- \}$ for scalars while $\{\eta_+, \,\bar\eta_+\}$ and $\{\eta_-, \,\bar\eta_-\}$ for fermions.
The (Grassmannian) sources with subscripts $+$ and $-$ are required to generate $\frac{1}{p^2-m^2+i\,\epsilon}$ and
$\frac{1}{p^2-m^2-i\,\epsilon}$ respectively. We will carry out this strategy in what follows.

\subsection{Scalars}

For a non-interacting real scalar $\phi$ with mass $m$, we propose that the generating functional in the presence of sources
$\{J_+,\, J_- \}$ is given by
\bea
\label{functionalS}
Z[J_+, J_-] &=& \int\, \mathcal{D}\phi \,\, \exp\left\{\,\,i\,\int\,d^4x\,\left(\,\,\mathcal{L}(\phi) + \frac{1}{2}\,J_+(x)\,\phi(x)
+\frac{1}{2}\,J_-(x)\, \phi(x) \,\,\right)  \,\,\right\}  \\
\label{functionalS2}
&=& \int\, \mathcal{D}\phi \,\, \exp\left\{\,\,\frac{i}{2}\,\int\,d^4x\,\left(\,\,\mathcal{L}(\phi)
+ J_+(x)\,\phi(x) +\frac{i}{2}\,\epsilon\, \phi^2(x)\,\,\right)\,\,\right\}\,
\nonumber \\
&& ~~~~~~~~~\, \exp\left\{\,\,\frac{i}{2}\,\int\,d^4x\,\left(\,\,\mathcal{L}(\phi)
+ J_-(x)\,\phi(x) -\frac{i}{2}\,\epsilon\, \phi^2(x) \,\,\right)\,\,\right\} \,,
\eea
with $\epsilon \rightarrow 0^{+}$. At first glance, it is not obvious that the path integral above is convergent as the factors
$+\frac{i}{2}\,\epsilon\, \phi^2(x)$ and $-\frac{i}{2}\,\epsilon\, \phi^2(x)$ cancel each other. One way to justify the
convergence is to think of these two factors as $+\frac{i}{2}\,\epsilon_+\, \phi^2(x)$ and $-\frac{i}{2}\,\epsilon_-\, \phi^2(x)$
with $\epsilon_{\pm} \rightarrow 0^{+}$. The convergence of the path integral can then be guaranteed by assuming
$\epsilon_+ \, - \, \epsilon_-\, \rightarrow\, 0^{+}$. For simplicity, we will set $\epsilon_+ = \epsilon_- = \epsilon$ from now on.

Except from an extra overall factor of $1/2$, the first term in Eq. \eqref{functionalS2} is the usual path integral for
a non-interacting real scalar $\phi$. For the second term in Eq. \eqref{functionalS2},
one simply needs to replace $m^2 - i\,\epsilon $ by $m^2 + i\,\epsilon $ in the usual path integral for
a non-interacting real scalar $\phi$ (apart from the extra overall factor of $1/2$). Thus,
following the similar derivations as in the usual path-integral formulation, the normalized generating
functional will be given by
\bea
Z_0[J_+, J_-] = Z_0[J_+]\, Z_0[J_-]\,,
\eea
where
\bea
\label{functionalS0plus}
Z_0[J_+] &=& \exp\left(\,-\frac{i}{4}\,\int\,d^4x\,d^4y\,\, J_+(x) \, \Delta_+(x-y)\, J_+(y) \,\right)\,,~ \\
\label{functionalS0minus}
Z_0[J_-] &=& \exp\left(\, -\frac{i}{4}\,\int\,d^4x\,d^4y\,\, J_-(x) \, \Delta_-(x-y)\, J_-(y) \,\right)\,.~
\eea
The Green's functions $\Delta_+(x-y)$ and $\Delta_-(x-y)$:
\bea
\Delta_{+}(x-y) &=& \int\,\frac{d^4\,p}{(2\pi)^4}\,\, e^{-ip(x-y)}\, \frac{1}{p^2-m^2+i\,\epsilon}\,, ~~~\\
\Delta_{-}(x-y) &=& \int\,\frac{d^4\,p}{(2\pi)^4}\,\, e^{-ip(x-y)}\, \frac{1}{p^2-m^2-i\,\epsilon}\,, ~~~
\eea
obey the inhomogeneous equations $\left(\,\Box^2 + m^2-i\epsilon\,\right)\,\Delta_+(x-y) = - \delta^{(4)}(x-y)$ and $\left(\,\Box^2 + m^2+i\epsilon\,\right)\,\Delta_-(x-y) = - \delta^{(4)}(x-y)$ respectively.

The two-point correlation function can be obtained from
\bea
\langle 0| \,T\{\,\phi(x)\,\phi(y)\,\}\,| 0 \rangle
&=& \left. \frac{1}{i}\,\left(\frac{\delta}{\delta J_+(x)} + \frac{\delta}{\delta J_-(x)} \right) \,
\frac{1}{i}\,\left( \frac{\delta}{\delta J_+(y)} +\frac{\delta}{\delta J_-(y)}  \right) \, Z_0[J_+, J_-] \, \right|_{J_\pm = 0} ~~~~~~~~ \\
&=& \frac{1}{2}\,\left[\, i\,\Delta_+(x-y) + i\,\Delta_-(x-y)\, \right]\,,  \nonumber
\eea
which is consistent with Eq. \eqref{P_S}. This can be generalized to the $n$-point correlation function:
\bea
\label{correlationS}
&& \langle 0 | \,T \left\{ \,\phi(x_1)\,\cdots\,\phi(x_n)\,\right\} \,| 0 \rangle  \nonumber \\
&=& \left.\frac{1}{i^n} \, \left(\frac{\delta}{\delta J_+(x_1)} + \frac{\delta}{\delta J_-(x_1)} \right) \, \cdots \,
\left(\frac{\delta}{\delta J_+(x_n)} + \frac{\delta}{\delta J_-(x_n)} \right) \, \, Z_0[J_+, J_-]\, \right|_{J_{\pm}=0}\,.
\eea

For a non-interacting complex scalar, $\mathcal{D}\phi$ in Eq. \eqref{functionalS} is replaced by
$\mathcal{D}\phi^\dagger\,\mathcal{D}\phi$ and $J_\pm(x)\,\phi(x)$ by $J_\pm^\dagger(x)\,\phi(x) + \phi^\dagger(x)\,J_\pm(x) $.
In Eq. \eqref{functionalS2}, $\pm\frac{i}{2}\,\epsilon\, \phi^2(x)$ is replaced by $\pm \, i\,\epsilon\, \phi^\dagger(x)\,\phi(x)$. In
Eq. \eqref{functionalS0plus} and Eq. \eqref{functionalS0minus}, $J_\pm(x)$ is replaced by $J_\pm^\dagger(x)$ and
the overall factor is \,$-i/2$\, instead of \,$-i/4$. The $2n$-point correlation function is
\bea
\label{correlationComplexS}
&& \langle 0 | \, T \left\{ \,\phi(x_1)\,\phi^\dagger(x_2)\,\cdots\,\phi(x_{2n-1})\,\phi^\dagger(x_{2n})\,\right\} \,| 0 \rangle \nonumber \\
&=& \frac{1}{i^{2n}} \,\left(\frac{\delta}{\delta J^\dagger_+(x_1)} + \frac{\delta}{\delta J^\dagger_-(x_1)} \right)
\left(\frac{\delta}{\delta J_+(x_2)} + \frac{\delta}{\delta J_-(x_2)} \right) \, \cdots \, \nonumber \\
&& \left.\left(\frac{\delta}{\delta J^\dagger_+(x_{2n-1})} + \frac{\delta}{\delta J^\dagger_-(x_{2n-1})} \right)
\left(\frac{\delta}{\delta J_+(x_{2n})} + \frac{\delta}{\delta J_-(x_{2n})} \right) \,\,
Z_0[J_+, J_+^\dagger, J_-, J_-^\dagger]\, \right|_{J_{\pm}=J^\dagger_{\pm}= 0}\,. ~~~~~~
\eea

\subsection{Fermions}

For a non-interacting fermion $\psi$ with mass $m$, we propose that the generating functional in the presence of sources
$\{\eta_+, \,\bar\eta_+\}$ and $\{\eta_-, \,\bar\eta_-\}$ is given by
\bea
\label{functionalF}
&& Z[\eta_+, \bar\eta_+, \eta_-, \bar\eta_-]  \nonumber \\
&=& \int\, \mathcal{D}\bar\psi \,\mathcal{D} \psi \,\, \exp\left\{\,\,i\,\int\,d^4x\,
\left[\,\,\mathcal{L}(\bar\psi,\psi) + \frac{1}{2}\,\left(\,\bar\eta_+(x)\,\psi(x)+\bar\psi(x)\,\eta_+(x)\,\right) \right.\right.\nonumber \\
&& \left.\left. ~~~~~~~~~~~~~~~~~~~~~~~~~~~~~~~~~~~~~~~~~~~~~~\,
+\frac{1}{2}\,\left(\,\bar\eta_-(x)\,\psi(x)+\bar\psi(x)\,\eta_-(x)\,\right) \,\,\right] \,\,\right\}  \\
\label{functionalF2}
&=& \int\, \mathcal{D}\bar\psi \,\mathcal{D} \psi \,\, \exp\left\{\,\,\frac{i}{2}\,\int\,d^4x\,\left(\,\,\mathcal{L}(\bar\psi,\psi)
+ \bar\eta_+(x)\,\psi(x)+\bar\psi(x)\,\eta_+(x) + i\,\epsilon\,\bar\psi\,\psi\,\,\right)\,\,\right\}\,
\nonumber \\
&& ~~~~~~~~~~~~~~\, \exp\left\{\,\,\frac{i}{2}\,\int\,d^4x\,\left(\,\,\mathcal{L}(\bar\psi,\psi)
+ \bar\eta_-(x)\,\psi(x)+ \bar\psi(x)\,\eta_-(x)- i\,\epsilon\,\bar\psi\,\psi \,\,\right)\,\,\right\} \,,~~~
\eea
with $\epsilon \rightarrow 0^{+}$. Similar to the scalar case, it is not obvious that the path integral above is convergent as the factors
$+ i\,\epsilon\,\bar\psi\,\psi$ and $- i\,\epsilon\,\bar\psi\,\psi$ cancel each other. One way to justify the
convergence is to think of these two factors as $+ i\,\epsilon_+\,\bar\psi\,\psi$ and $- i\,\epsilon_-\,\bar\psi\,\psi$
with $\epsilon_{\pm} \rightarrow 0^{+}$. The convergence of the path integral can then be guaranteed by assuming
$\epsilon_+ \, - \, \epsilon_-\, \rightarrow\, 0^{+}$. For simplicity, we will set $\epsilon_+ = \epsilon_- = \epsilon$ from now on.

Except from an extra overall factor of $1/2$, the first term in Eq. \eqref{functionalF2} is the usual path integral for
a non-interacting fermion $\psi$. For the second term in Eq. \eqref{functionalF2},
one simply needs to replace $m^2 - i\,\epsilon $ by $m^2 + i\,\epsilon $ in the usual path integral for
a non-interacting fermion $\psi$ (apart from the extra overall factor of $1/2$). Thus,
following the similar derivations as in the usual path-integral formulation, the normalized generating
functional will be given by
\bea
Z_0[\eta_+, \bar\eta_+, \eta_-, \bar\eta_-] = Z_0[\eta_+, \bar\eta_+]\, Z_0[\eta_-, \bar\eta_-]\,,
\eea
where
\bea
Z_0[\eta_+, \bar\eta_+] &=& \exp\left(\,-\frac{i}{2}\,\int\,d^4x\,d^4y\,\, \bar\eta_+(x) \, \mathcal{S}_+(x-y)\, \eta_+(y) \,\right)\,, \nonumber \\
Z_0[\eta_-, \bar\eta_-] &=& \exp\left(\,-\frac{i}{2}\,\int\,d^4x\,d^4y\,\, \bar\eta_-(x) \, \mathcal{S}_-(x-y)\, \eta_-(y) \,\right)\,.
\eea
The Green's functions $\mathcal{S}_+(x-y)$ and $\mathcal{S}_-(x-y)$
\bea
\mathcal{S}_+(x-y) &=& \int\,\frac{d^4 p}{(2\pi)^4}\, e^{-ip(x-y)} \,\frac{\slashed{p}+m}{p^2-m^2 +i\epsilon }\,, \nonumber \\
\mathcal{S}_-(x-y) &=& \int\,\frac{d^4 p}{(2\pi)^4}\,e^{-ip(x-y)} \,\frac{\slashed{p}+m}{p^2-m^2 -i\epsilon } \,.
\eea
obey the inhomogeneous equations $\left(\,\Box^2 + m^2-i\epsilon\,\right)\,\mathcal{S}_+(x-y) = - \delta^{(4)}(x-y)$ and $\left(\,\Box^2 + m^2+i\epsilon\,\right)\,\mathcal{S}_-(x-y) = - \delta^{(4)}(x-y)$ respectively.

The two-point correlation function can be obtained from
\bea
&& \langle 0| \,T\{\,\psi(x)\,\bar\psi(y)\,\}\,| 0 \rangle \nonumber \\
&=& \left. \frac{1}{i}\,\left(\frac{\delta}{\delta\bar \eta_+(x)} +\frac{\delta}{\delta \bar\eta_-(x)} \right) \,
\left(-\frac{1}{i}\right)\,\left( \frac{\delta}{\delta \eta_+(y)} + \frac{\delta}{\delta \eta_-(y)}  \right) \, Z_0[\eta_+,\bar\eta_+, \eta_-,\bar\eta_-] \, \right|_{\eta_\pm =\bar\eta_\pm = 0} ~~~~ \\
&=& \frac{1}{2}\,\left[ \,i \,\mathcal{S}_+(x-y) + i\,\mathcal{S}_-(x-y)\, \right]\,,
\eea
which is consistent with Eq. \eqref{P_F}. This can be generalized to the $2n$-point correlation function:
\bea
&& \langle 0 | \,T \left\{ \,\psi(x_1)\bar\psi(x_2)\,\cdots\,\psi(x_{2n-1})\bar\psi(x_{2n})\,\right\} \,| 0 \rangle \nonumber \\
&=& \left(\frac{\delta}{\delta \bar\eta_+(x_1)} +\frac{\delta}{\delta \bar\eta_-(x_1)} \right) \,
\left( \frac{\delta}{\delta \eta_+(x_2)} + \frac{\delta}{\delta \eta_-(x_2)}  \right)\, \cdots \nonumber \\
&&  \left. \left(\frac{\delta}{\delta \bar\eta_+(x_{2n-1})} + \frac{\delta}{\delta \bar \eta_-(x_{2n-1})} \right) \,
\left( \frac{\delta}{\delta \eta_+(x_{2n})} + \frac{\delta}{\delta \eta_-(x_{2n})}  \right)\,\,
Z_0[\eta_+,\bar\eta_+, \eta_-,\bar\eta_-] \, \right|_{\eta_\pm =\bar\eta_\pm = 0}\,. ~~~~~~~~
\eea

\subsection{Interacting Theory}

When interactions are present, the most general generating functional involving a scalar $\phi$, a fermion $\psi$, an abelian vector boson
$A_\mu$ and a nonabelian vector boson $A^a_\nu$ is given by
\bea
Z[J_+, J_+^\dagger, J_-, J_-^\dagger, \eta_+, \bar\eta_+, \eta_-, \bar\eta_-, J_\mu, J^a_\nu] = \exp\left(\,i\,\int\, d^4x\,\,\tilde{\mathcal{L}}_I \,\right)\, \tilde{Z}_{\textrm{0}}\,,
\eea
where $\tilde{Z}_{0}$ is the product of the normalized generating functionals for the scalar, fermion, abelian and nonabelian
vector bosons. $\tilde{\mathcal{L}}_I$ is the Lagrangian that contains the supersymmetric interaction operators under consideration, but
with the fields for the scalar, fermion, abelian and nonabelian vector bosons substituted by:
\bea
\phi(x) ~ &\rightarrow& ~ \frac{1}{i}\,\left(\frac{\delta}{\delta J_+(x)} + \frac{\delta}{\delta J_-(x)} \right)~,
~~~~~~\textrm{for\, real}~ \phi \\
\phi(x) ~ &\rightarrow& ~ \frac{1}{i}\,\left(\frac{\delta}{\delta J^\dagger_+(x)} + \frac{\delta}{\delta J^\dagger_-(x)} \right)~,
~~~~~~\textrm{for\, complex}~ \phi \\
\phi^\dagger(x) ~ &\rightarrow& ~ \frac{1}{i}\,\left(\frac{\delta}{\delta J_+(x)} + \frac{\delta}{\delta J_-(x)} \right)~,
~~~~~~\textrm{for\, complex}~ \phi \\
\psi(x) ~ &\rightarrow& ~ \frac{1}{i}\,\left(\frac{\delta}{\delta \bar\eta_+(x)} + \frac{\delta}{\delta \bar\eta_-(x)} \right)~, \\
\bar\psi(x) ~ &\rightarrow& ~ -\frac{1}{i}\,\left(\frac{\delta}{\delta \eta_+(x)} + \frac{\delta}{\delta \eta_-(x)} \right)~, \\
A_\mu(x) ~ &\rightarrow& ~ \frac{1}{i}\,\frac{\delta}{\delta J^\mu}~, ~~~~~~
A^a_\mu(x) ~ \rightarrow ~ \frac{1}{i}\,\frac{\delta}{\delta J_a^\mu}~.
\eea

\section{Conclusions}

The idea of off-shell supersymmetry evades the direct detection bounds on sparticles. On the other hand, it
solves the gauge hierarchy problem and unifies the gauge couplings in the usual way. This idea fundamentally changes our
understanding about supersymmetry. It explains why we have not directly observed the sparticles at the LHC so far and suggests a new
way of searching for them.

The original realization of off-shell supersymmetry relies on a generalization of the way we quantize the quantum fields and its
sole application to sparticles. A natural way to reinforce its consistency is through
the path-integral approach. In this paper, we show that off-shell supersymmetry can be formulated from the path-integral approach. \\


\newpage

\emph{Acknowledgments.}\,\, We thank Pierre Ramond for useful conversations.
CMH was supported by the Office of the Vice-President for Research and Graduate
Studies at Michigan State University.


\end{document}